# Exceed Improved Efficient Perovskite Solar Cells Under Dual-Irradiation System

Tao Ye[1,2†], Xianqiang Li[1,3†], Shaoyang Ma[3], Dan Wu[3], Lei Wei[3], Xiaohong Tang[3], Jian Wei Xu[1*], Seeram Ramakrishna[2], Chellappan Vijila[1*], Xizu Wang[1*†]

[1]Institute of Materials Research and Engineering (IMRE), Agency for Science, Technology and Research (A*STAR), #08-03, 2 Fusionopolis Way, Innovis, 138634, Singapore

[2]Department of Mechanical Engineering, National University of Singapore, Singapore 117576, Singapore

[3]School of Electrical and Electronic Engineering, Nanyang Technological University, 50 Nanyang Avenue, Singapore 639798, Singapore

†These authors contributed equally to this work.

*Email: jw-xu@imre.a-star.edu.sg (J.W.X.); c-vijila@imre.a-star.edu.sg (C.V.); wangxz@imre.a-star.edu.sg (X.W.)

**Abstract**: In general, perovskite solar cells (PSC) with a sensitized or thin film architecture absorb light from a single illumination. This paper reported a PSC architecture with a semitransparent Au/ITO counter electrode, which allows light to pass it partially. When the device was illuminated simultaneously from both the FTO and Au/ITO sides, the PSC has achieved an overall power conversion efficiency (PCE) as high as 20.1% under high light intensity (1.4 sun), which is much higher than that of the single-irradiation system.

The lead halide perovskite $APbX_3$ [A = Cs (cesium), MA ($CH_3NH_3$, methylammonium), or FA ($NH=CHNH_3$, formamidinium); X = Cl, Br and I] solar cells are a strong competitor to the traditional silicon solar cells and III–V solar cells for the future clean energy applications due to the unique optoelectronic properties of perovskites, such as optical tunability, high absorption coefficients, millimeter-scale charge carriers diffusion length as well as easy fabrication[1-11]. The latest highest certificated power conversion efficiency (PCE) of the perovskite solar cell (PSC) is 22.1% and this PCE is achieved with a mixed perovskite light absorber fabricated through a two-step intramolecular exchanging process[1,8]. In contrast, the PSC fabricated with one step deposition of mixed perovskite light absorber [$FA_{0.81}MA_{0.15}Pb(I_{0.836}Br_{0.15})_3$] can achieve a comparable PCE (21.6%)[9-13] to the highest certificated one. Tremendous attention has been paid to the mixed perovskites, which are fabricated through mixing cations and halides since pure $APbI_3$-based perovskites show numerous disadvantages in solar cell fabrication, for example, unfavorable structural phase transition and light-induced trap-state ($MAPbI_3$), degradation upon contact with moisture and solvents ($MAPbI_3$ and $FAPbI_3$), the formation of photo-inactive hexagonal δ-phase and photoactive α-phase during crystallization ($FAPbI_3$), and high temperature to form photo-active phase

($CsPbI_3$)[8,14-22]. Mixed perovskite systems have shown the integrated advantages of the constituents while avoiding the weaknesses[8-11,23].

Some transparent electrodes to allow the light to go into the device from both electrode sides have been developed, and this has been discussed in some early photo-electrochemical devices[24-26]. However, in all the existing PSCs even including all types of other photovoltaic devices, light always goes into the device from the transparent conductive electrode side and the incident light only comes from a single light source. The rear metal electrode mainly functions as a reflector to generate additional energy from diffused light and reflected light to enhance to some extent the absorption efficiency. In this study, unlike a traditional PSC with a thick real metal electrode, the thickness of the rear metal electrode is reduced to 15-30 nm from 60 to 120 nm, and transparent indium tin oxide (ITO) is deposited on the thin Au layer to form a semitransparent Au/ITO electrode[27], which will allow more light to pass through the rear layer. A dual-irradiation PSC system allowing light to enter into the system from both sides gave the PCE of 20.1 ± 0.8% under dual-irradiation [simulated 1 sun (AM 1.5G) at the FTO side and a white LED at the Au/ITO side]. The new PSC structures with a dual-irradiation system offer a new insight into perovskite photovoltaic physics and device integration, and this type of novel PSC dual-irradiation system is potentially useful for many applications.

**Preparation of the semitransparent Au/ITO electrode**

The dual-irradiation device structure with a configuration of FTO/Cl-TiO$_2$/Mp-TiO$_2$/mixed perovskite/spiro-OMeTAD/Au/ITO, and the energy levels[12,23,28] of different device components are shown in Figure 1a. When light enters the device, the perovskite layer absorbs light and generates free holes and electrons. The free electrons transport through the TiO$_2$ layers to the FTO electrode, while the free holes are extracted through the spiro-OMeTAD layer and collected by the Au/ITO electrode. The device fabrication details are given in the Methods section. The top Au/ITO electrode is optimized to enhance the light harvesting of the dual irradiation system. The fabricated ITO layer properties have been investigated through scanning electron microscope (SEM), atomic force microscope (AFM), X-ray photoelectron spectroscopy (XPS) and ultraviolet photoelectron spectroscopy (UPS) methods, as shown in Figure 2. A compact and smooth ITO surface can be seen in Figure 2a. And the roughness of the ITO layer is investigated through AFM and the average roughness (Ra) is just 2.5 nm for this sample (Figure 2b). XPS measurement was conducted to explore atomic and electronic structures of the prepared ITO layer. Figure 2c, 2d and 2e show the XPS elemental spectra (In 3d, Sn 3d and O 1s) of the sample. According to some recent publications[24,29-31], the binding energy peaks of the XPS elemental spectra show that the ITO precursor was successfully transformed into ITO in this study. UPS has been used for the work function measurement of the prepared ITO layer in this study. As explicated in Figure 2f, the work function of the prepared ITO layer is ~4.6 eV, which is similar to some recent reported ITO materials used as various transparent conductive electrode[29-31]. The cross-section SEM image of the device is shown in Figure 1b, which illustrates the thickness and morphology of each functional layers of the device and the ITO is well deposited on top of the thin film Au.

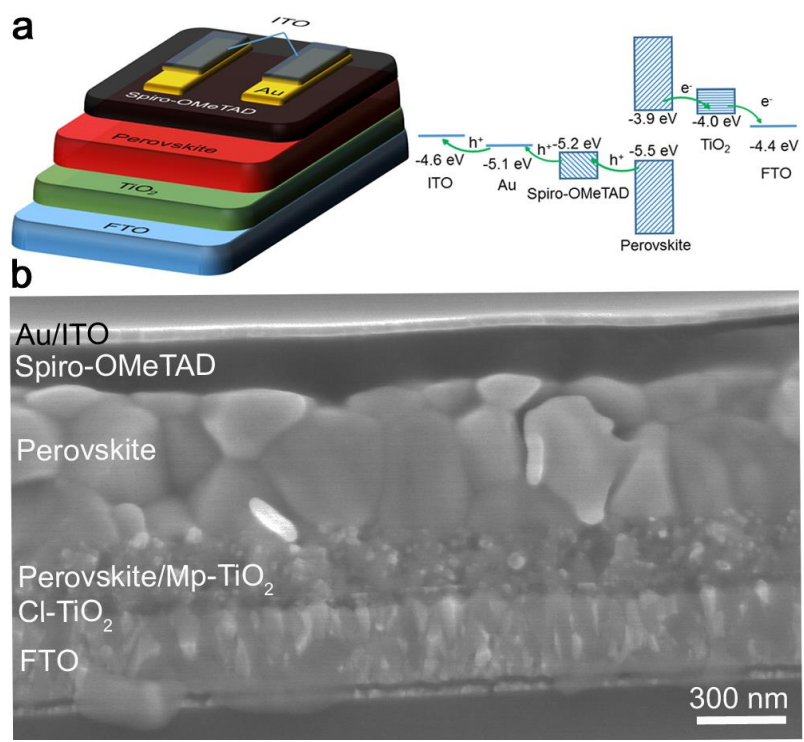

**Figure 1. Fundamentals of the dual-irradiation system.** (**a**) Left: schematic drawing of the dual-irradiation system; Right: the energy level diagram of the device; (**b**) The cross-section SEM image of the fabricated device with a 22.5 nm Au layer. Cl-TiO$_2$: compact TiO$_2$; Mp-TiO$_2$: mesoporous TiO$_2$; spiro-OMeTAD: 2,2',7,7'-Tetrakis[N,N-di(4-methoxyphenyl)amino]-9,9'-spirobifluorene.

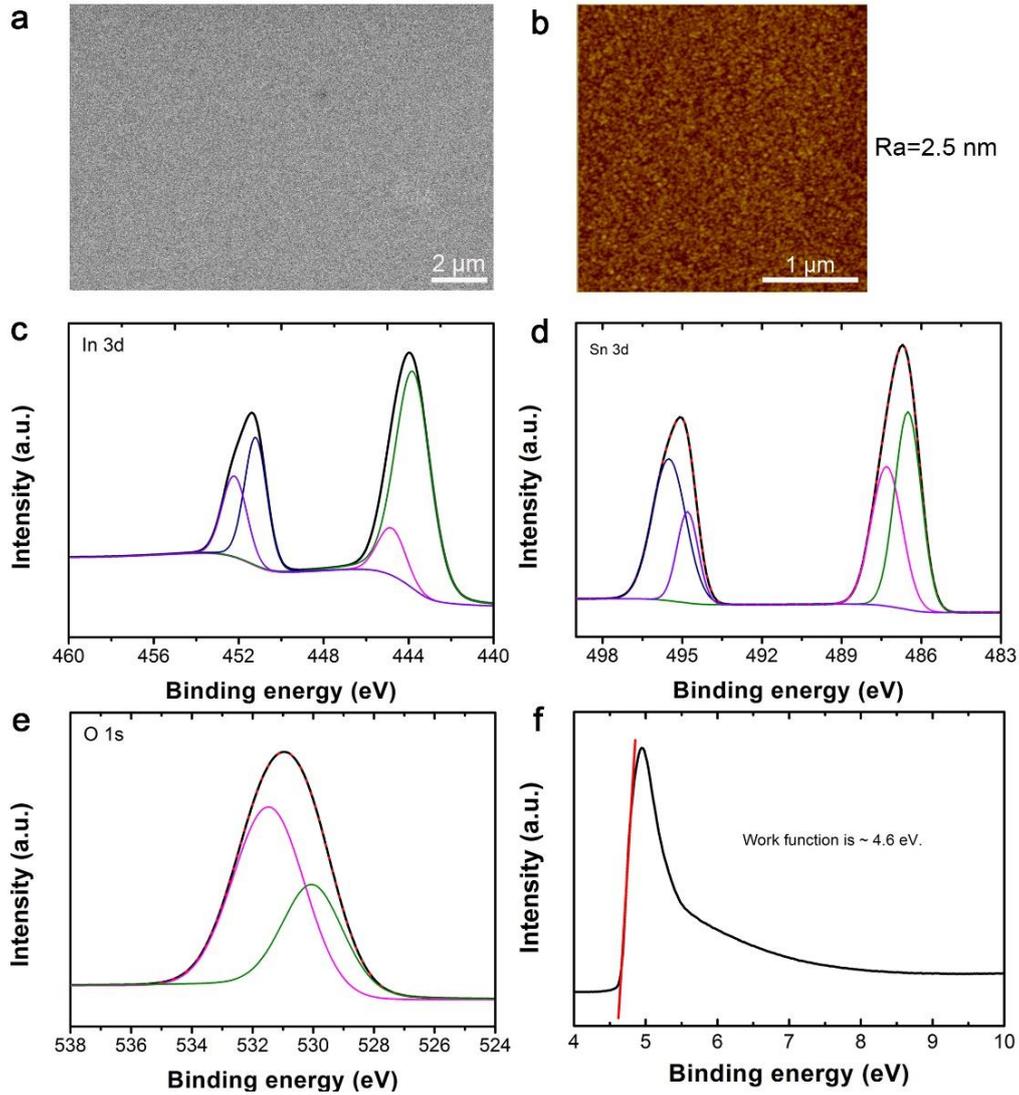

**Figure 2.** Fundamentals of prepared ITO layer. (**a**) Surface SEM image shows the smooth and uniform prepared ITO layer. (**b**) AFM image of the prepared ITO layer. (**c-e**) XPS elemental spectra (In 3d, Sn 3d and O 1s) of the ITO sample. (**f**) UPS work function spectra of the fabricated ITO layer.

**Single-side illumination performance**

Current density-voltage (*J-V*) curves and corresponding IPCE spectra of one side illumination of the different PSCs are shown in Figure 3. According to the *J-V* results, we can see that the performance of the devices with semi-transparent ITO electrodes are depended on the thickness of the Au electrodes. As for the device with the thinnest (17.5 nm) Au electrode, the device performance measured from FTO side is poor due to the insufficient free holes collection of the thin Au layer, however, the thin Au layer can let more light in from the ITO side so the highest PCE is observed within the same device. When the thickness of the Au electrode increasing in the device, nearly all the photovoltaic (PV) parameters for the FTO side are enhanced (as can be seen from Fig 2a and 2b) but all PV parameters are going down for the ITO side [especially for short circuit current density ($J_{sc}$), as can be seen from Figure 3c]. These results can be further verified by the IPCE results, the IPCE results from FTO and ITO sides of the PV devices can be found in Figure 3b and 3d, respectively. More precisely, the IPCE shapes and values of the FTO side are similar in this study (or only slightly increased when the thickness of the Au electrode increased from 17.5 nm to 27.5 nm). However, the IPCE is enhanced at a wide wavelength range of 400 ~ 800 nm (especially for 600 ~ 800 nm due to the light attenuation of Au electrode, as can be seen from Figure 4) when the Au electrode decreased from 27.5 nm to 17.5 nm. Thus, the device performance from FTO and Au/ITO side can be adjusted by controlling the thickness of Au electrode.

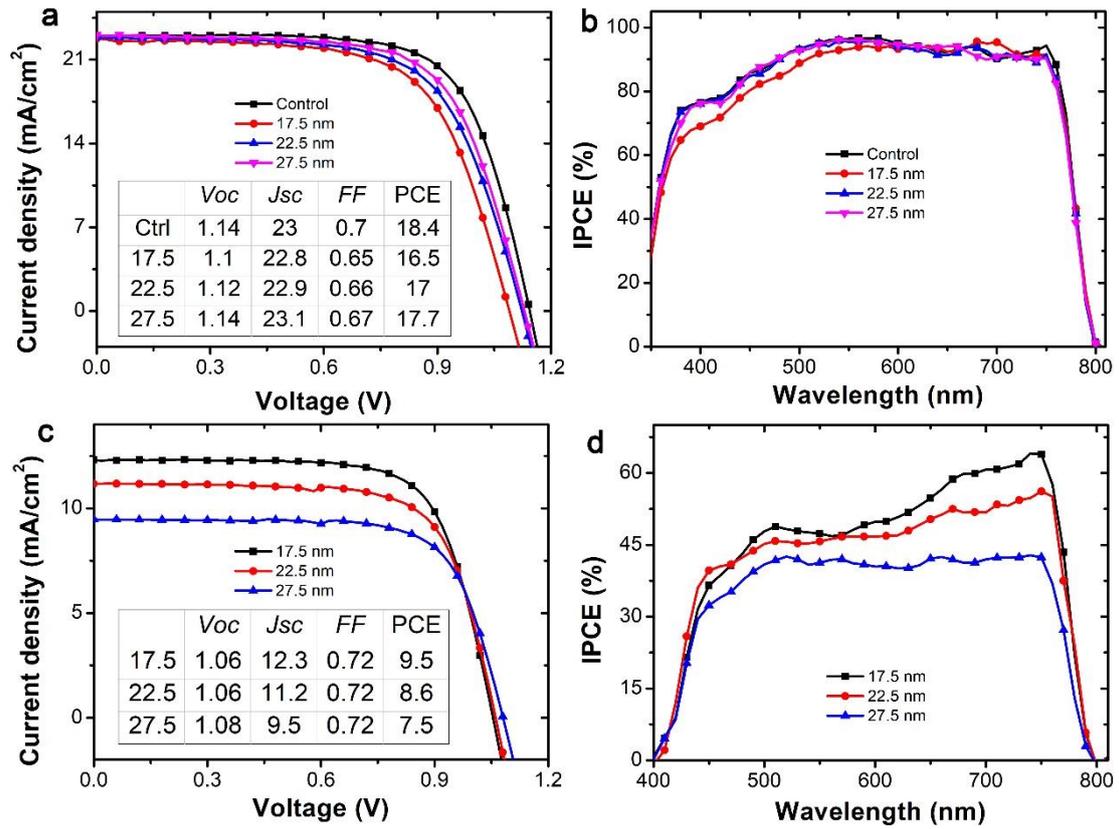

**Figure 3. Single-irradiation photovoltaic parameters.** (**a**) & (**b**) *J-V* curve and IPCE of different devices obtained from FTO sides; (**c**) & (**d**) *J-V* curve and IPCE of different devices obtained from ITO sides. Control group: device with 80 nm Au single electrode.

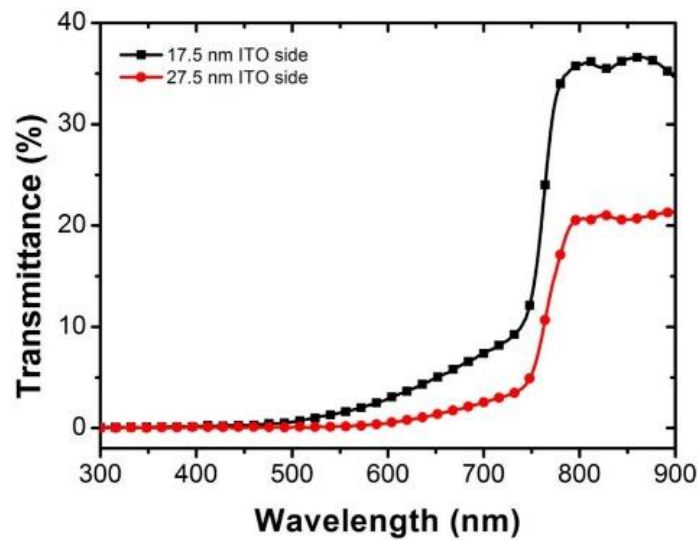

**Figure 4**. Optical transmittance of the PSCs based on 17.5 and 27.5 nm Au electrodes, the measuring lights goes in from ITO side. With the thickness of the Au layer

increase, the optical transmittance from ITO side is decreased due to the light attenuation of Au electrode. And we can see that the optical transmittance start point is ~500 nm of the 17.5 nm Au device and it is ~600 nm for 27.5 nm Au device.

The *J-V* curves of the PSC device based on different irradiation conditions have been collected under a wide simulated light intensity range (from 0.1 to 1.4 sun, Figure 5 and 6). As for FTO side, the open circuit voltage ($V_{oc}$) of the device is not remarkable changed but the fill factor (*FF*) and $J_{sc}$ of the device show significant changes. Specifically, the *FF* decrease due to the increased free carrier recombination derived from the rapid increase of the free carrier density when the light intensity increases. However, the increased free carrier density also leads to the $J_{sc}$ improvement of the device (Figure 5b). As for the ITO side, we observed the same trend of the PV parameters for the same device (Figure 6). However, because of the light attenuation of the Au electrode, the carrier density of the ITO side is lower than that in the FTO side, the *FF* decrease is not significant when the *J-V* curves are measured from ITO side below 1 sun. Thus, the $J_{sc}$ from both FTO and ITO sides show a linear relationship (Figure 5b and 6b) with the simulated light intensity, indicating the charge carriers recombination caused by defects is low in the device[32-34].

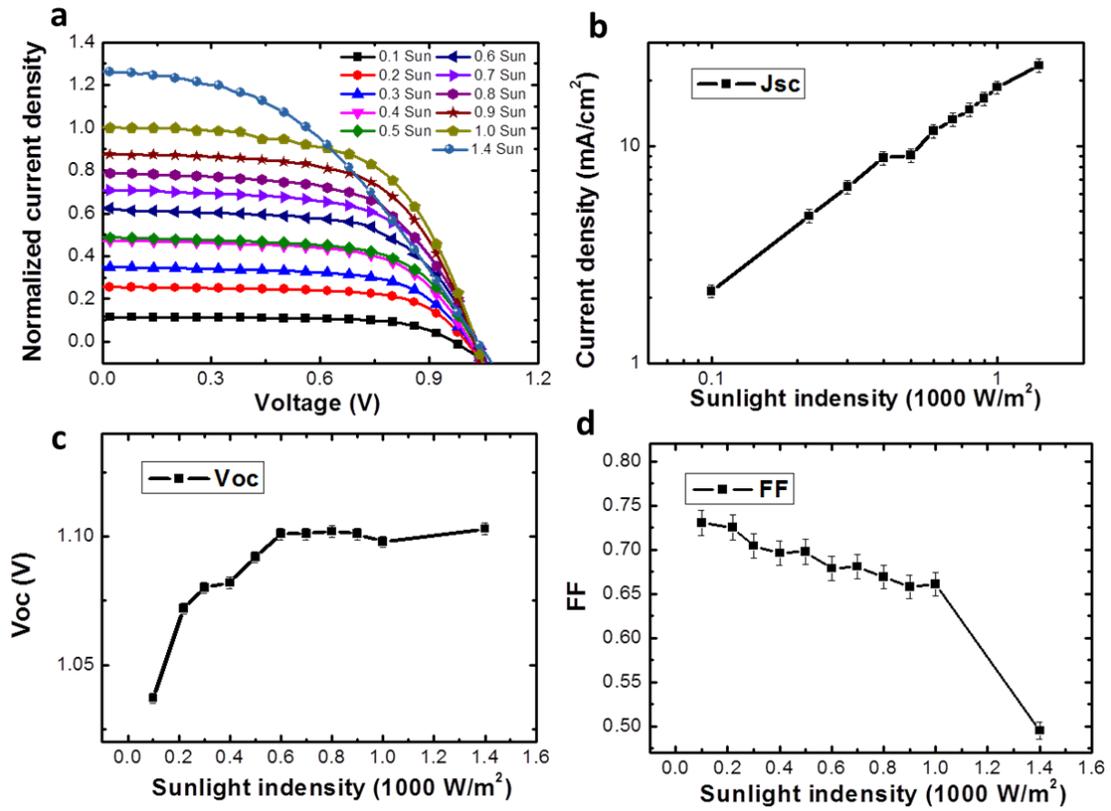

**Figure 5.** (**a**) Light intensity response of the device performance for an Au-22.5 device when the light goes in from the FTO side; (**b**) the log-log plot of the $J_{sc}$ against the light intensity; (**c**) the $V_{oc}$ and (**d**) FF against the light intensity.

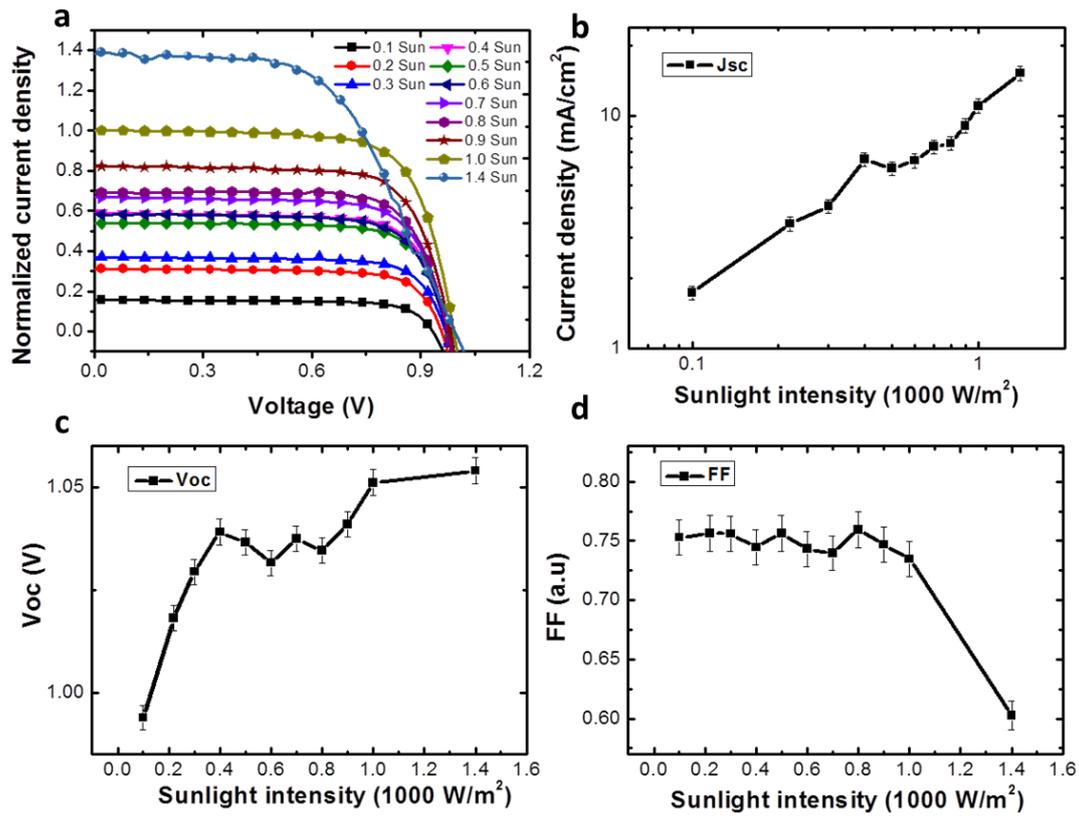

**Figure 6.** (**a**) Light intensity response of the device performance for an Au-22.5 device when the light goes in from the ITO side; (**b**) the log-log plot of the $J_{sc}$ against the light intensity; (**c**) the $V_{oc}$ and (**d**) $FF$ against the light intensity.

**Dual-side illumination performance**

A white LED has been utilized for providing another light source for the performance measurement of the same device and the full light spectrum of the LED can be found in Figure 7. Figure 8a schematically shows the testing system of the device under dual-irradiation conditions. The $J$-$V$ curves of the PSC device based on different irradiation conditions are given in Figure 8b and the data are summarized in Table 1. More specifically, the ITO side shows an overall PCE of 16.6% with a $V_{oc}$ of 1.06 V, $J_{sc}$ of 98 mA/cm$^2$, and $FF$ of 0.64 under the LED illumination, meanwhile the FTO side shows an overall PCE of 16.5% with a $V_{oc}$ of 1.10V, $J_{sc}$ of 22.6 mA/cm², and *FF* of 0.67 under the simulated 1 sun illumination. Here is a simple addition: the ITO LED *J-V* curve pluses the FTO solar simulator *J-V* curve in the numerical addition, an overall PCE of 16.2% is achieved at 1.4 sun. It is surprising that when the device works under dual irradiation method at same time (LED light goes in from ITO side and 1 sun light goes in from FTO side), an enhanced PCE over 20.1% has been observed with a $V_{oc}$ of 1.08 V, $J_{sc}$ of 40.4 mA/cm², and *FF* of 0.64. Thus, overall device performance enhancement has been achieved for the device when the white LED is utilized as another light source and lights on the ITO side of the device. The PCE is much higher than that of both one side irradiations under high light intensity (1.4 sun) conditions, and also higher than the sum of the single side irradiation. This inspiring improvement of PCE under the dual-irradiation system impels us to investigate its optical/electronic conversion mechanism behind.

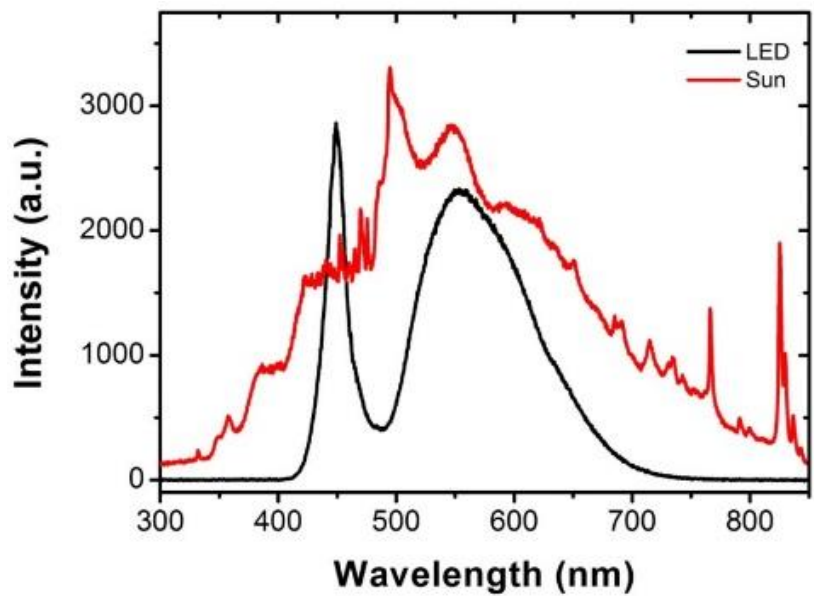

**Figure 7.** The light spectra of white LED and simulated 1 sun light.

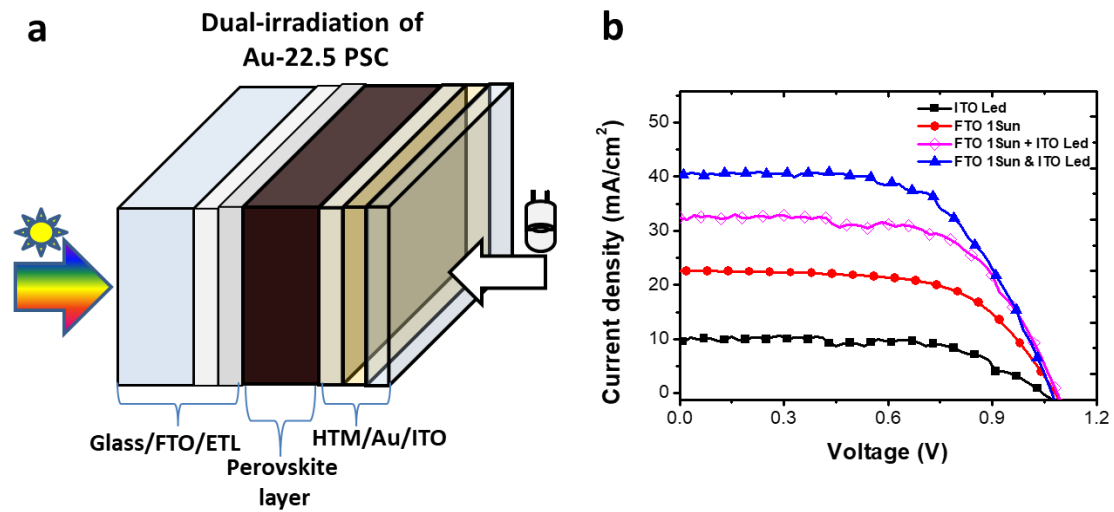

**Figure 8. Dual-irradiation photovoltaic parameters of semitransparent devices.** (**a**) schematic image of the dual-irradiation testing setup; (**b**) J-V curve of devices with the Au-22.5nm/ITO under one or two light sources single or dual irradiation. (LED goes in from ITO side and 1 sun simulated light goes in from FTO side). ETL: electron transport layer; HTM: hole transport material.

**Table 1. Photovoltaic parameters for the Au-22.5 device measured under different light intensity conditions.** All the PV parameters are calculated at specific light intensity. The PCE with error is avegrage from 12-20 unit devices of each experiment. ‡The sum of J-V curves of ITO LED + FTO solar simulator under the sum of two irradiation light intensity, 1400 W/m².

| Lighting source | Irradiation side | $V_{oc}$ (V) | $J_{sc}$ (mA/cm²) | FF | PCE (%) |
|---|---|---|---|---|---|
| LED* | ITO | 1.06 | 9.8 | 0.64 | 16.6 ± 0.8 |
| Solar Simulator* | FTO | 1.10 | 22.6 | 0.67 | 16.5 ± 0.5 |
| Solar Simulator** | FTO | 1.09 | 31.1 | 0.45 | 10.9 ± 0.3 |
| LED +Solar Simulator* | Dual-irradiation (ITO LED & FTO solar) | 1.08 | 40.4 | 0.64 | 20.1 ± 0.8 |

*light intensity solar simulator (1000 W/m²), LED (400 W/m²)

** light intensity solar simulator (1400 W/m²),

In conclusion, we have successfully fabricated highly efficient PSCs with a configuration of FTO/Cl-TiO$_2$/Mp-TiO$_2$/mixed perovskite/hole transport material/Au/ITO under a dual-irradiation system. The device performance is much better when it works under a dual-irradiation system than under a single-irradiation system. The device that absorbs light from both sides would open a new way for enhancing PCE to a higher level than its traditional counterparts. In addition, this type of dual-irradiation PSC systems could be developed with higher performance by modulating the optical and electrical characteristics of the rear electrode, and in turn their applicability can be extended in many other areas such as a substitute for glass walls/roofs in modern buildings, photo-detectors and sensors.

**Methods**

**Device fabrication**

FTO glass was cleaned with detergent, DI water, acetone (Sigma-Aldrich), isopropanol (Sigma-Aldrich) and treated with UV ozone treatment at 100 °C for 10 min. A ~30 nm $TiO_2$ layer was deposited on top of the FTO by spin-coating $TiO_2$ precursor solution at 6000 rpm for 30 s. Then, it was heated at 450 °C for 30 min in air. To prepare the $TiO_2$ precursor solution, titanium isopropoxide (1 mL, Sigma-Aldrich) and 12 M HCl solution (10 μL, Sigma-Aldrich) were diluted in ethanol (10 mL). Then a ~ 180 nm mesoporous $TiO_2$ layer was deposited by spin-coating a 30-nm $TiO_2$ nanoparticle paste (Dyesol) in ethanol (1:5.5 in weight ration) at a speed of 6000 rpm for 30 s. The substrate was then annealing at 500°C for 20 min. The $TiO_2$ layers were treated with a diluted $TiCl_4$ (50 mM in water, Sigma-Aldrich) solution[7]. Next, perovskite film was deposited on the substrate by a spin-coating process in glovebox, with 2000 rpm for 10 s, followed by 6000 rpm for 45 s. Then chlorobenzene (110 μL, Sigma-Aldrich) was dropped in 8-10 s during the second spin-coating process. The substrate was then heated at 100 °C, 60 min. The precursor solution was prepared by dissolving 0.265 g PbI$_2$ (TCI), 0.037 g PbBr$_2$ (Sigma-Aldrich), 0.011 g MABr and 0.094 g FAI (Dyesol) in 0.4 mL anhydrous N,N-dimethylformamide (Sigma-Aldrich) and 0.1 mL anhydrous dimethylsulfoxide (Sigma-Aldrich). Spiro-OMeTAD (Merck) was further deposited by spin-coating at 3000 rpm for 30 s. The spiro-OMeTAD solution was prepared by dissolving 74 mg spiro-OMeTAD, 28.5 μL 4-tert-butylpyridine (Sigma-Aldrich), 17.5 μL of a stock solution of 520 mg/mL lithium bis (trifluoromethylsulphonyl) imide (Sigma-Aldrich) in acetonitrile (Sigma-Aldrich) and 29 μL of a stock solution of 100 mg/mL tris(2-(1H-pyrazol-1-yl)-4-tert-butylpyridine)-cobalt(III) tris(bis(trifluoromethylsulfonyl)imide) (Sigma-Aldrich) in acetonitrile in 1 mL anhydrous chlorobenzene. Finally, 80 nm of gold was deposited as an electrode by thermal evaporation. As for the devices with semitransparent Au/ITO electrode, thin gold electrodes (17.5 nm, 22.5 nm and 27.5 nm) was deposited on top of the substrates, followed by an ITO transparent conductor. An oxidized target (6 inch in diameter) with In$_2$O$_3$ (Sigma-Aldrich) and SnO$_2$ (Sigma-Aldrich) in a weight ratio of 9:1 was employed for the deposition of the ITO electrode. The ITO electrode was prepared by DC magnetron sputtering with a power of 10 W, and the deposition rate was estimated to be about 2 nm/min. The base pressure in the sputtering system was ~2.0 × 10$^{-4}$ Pa. During the film deposition, an argon–hydrogen gas mixture was employed. The argon partial pressure was set at ~2.9 × 10$^{-1}$ Pa, the hydrogen partial pressure varied from 1.1 × 10$^{-3}$ to 4.0 × 10$^{-3}$ Pa to modulate and optimize the properties of ITO films.

**Characterization**

*J-V* characteristics were measured in the glovebox with a solar simulator (SAN-EI Electric XES-301S 300W Xe Lamp JIS Class AAA) and a Keithley 2400 sourcemeter. Another light source was generated from a LED during the *J-V* measurement (the full spectra of the LED can be seen in Figure 7). Various neutral density filters (Newport) were used to adjust the light intensity within this study. The solar cells were masked with metal apertures to define the active areas which were typically 0.09 cm$^2$. IPCE was recorded with a Keithley 2400 sourcemeter combined with an Oriel 300-W Xe lamp, an Oriel Cornerstone 130 monochromator and a SRS 810 lock-in amplifier (Stanford Research Systems). The light intensity was determined by a monosilicon detector (with KG-5 visible color filter) calibrated by the National Renewable Energy Laboratory to minimize spectral mismatch. The surface and cross section morphologies of the samples were investigated by a SEM (JEOL JSM-7001F) at 10 kV. Transmittance spectra were measured on a UV-VIS-NIR spectrophotometer (UV-3600, SHIMADZU). ICON-PKG atomic force microscopy (AFM) from Bruker was used to investigate the morphological influence of the ITO/Au on the active layer. The X-ray photoelectron spectroscopy (XPS) spectra were collected with a VG ESCALAB 220I-XL system equipped with an Al K α X-ray source (1486.6 eV). The excitation source of UPS was He I (hv = 21.2 eV). Photoelectrons were collected by a rotatable hemispherical electron energy analyzer. UPS spectra were measured in normal emission to examine shifts of the secondary electron edge.

**Acknowledgments**

T.Y. thanks the National University of Singapore for his research scholarship.

**Author contributions**

T.Y. and X.W. conceived the experiments; T.Y., X.L., D.W., S.M. and X.W. carried out the device fabrication, characterizations and performance measurements. All authors analyzed data and wrote the manuscript.

**Competing financial interests**

The authors declare no competing financial interests.